\documentstyle[11pt,Cargesepasp,twoside,epsf]{article}
\def\edcomment#1{\iffalse\marginpar{\raggedright\sl#1\/}\else\relax\fi}
\reversemarginpar

\begin{document}
\title{The Orion Star-Forming Region}
 \author{Lynne A. Hillenbrand \& John M. Carpenter}
\affil{California Institute of Technology; Pasadena CA 91125; USA}
 \author{Eric D. Feigelson}
\affil{Pennsylvania State University; University Park, PA, 16802; USA}

\begin{abstract}
General properties of the Orion star-forming region are discussed,
with a focus on the dense Orion Nebula Cluster (ONC). 
This cluster contains between 2500 and 4500 objects 
located within a few parsecs of the eponymous Trapezium stars. 
Its members are aged $<$1 to a few Myr and encompass the full spectrum 
of stellar masses $<$50 M$_\odot$, as well as brown dwarfs detected 
with masses as low as $<$0.02 M$_\odot$ (20 M$_{Jupiter}$) thusfar. 
Recent results from optical, near-infrared, and x-ray studies
of the stellar/sub-stellar population associated with this cluster are summarized.  
\end{abstract}

\section{Introduction}

``Orion'' is of course a rather large and daunting topic to which I can not do 
justice in a single talk.  For ample review material I direct your attention 
to the recent volume on ``The Orion Complex, Revisted'' 
edited by McCaughrean, Burkert, \& O'Dell
(2001, based on a conference held in the summer of 1997).

By way of introduction, let me remind folks that
the constellation of Orion is unique in that all of its constituent stars
are located at approximately the same distance from the Sun (350-500 pc).
This region of the sky is pervaded by a Giant Molecular Cloud usually thought of
as two entities, Orion A and Orion B, which together contain 1-3$\times10^4$ 
stars aged $<$1-3 Myr as well as various off-cloud populations aged 3-30 Myr 
old, including an OB association just to the west which is
part of the Gould's Belt system.  
The spatial extent of current and recent star formation
activity is more than 60 degree$^2$ on the sky. However, 
the stellar population over this area is still largely uncatalogued.
Relative proximity combined with projection
$\sim$15-25 degrees out of the plane of the Galaxy means that 
full census information can be obtained without
much of the ambiguity that plagues other similarly interesting 
star-forming regions located further away and closer to the Galactic plane.
The Orion complex contains the nearest Giant Molecular Cloud 
and also the nearest example of recent massive star formation; as such, 
it has been the subject of intense study over the past several decades.  
It continues to be a target of almost every new instrument or technology.

I will spend the majority of my time here discussing results obtained over
the past several years on the stellar population of the Orion Nebula Cluster 
Region.  Older results
include those on the Hertzsprung-Russell Diagram for an unbiased sample of
$\sim$1000 stars, the initial mass function, mass segregation,
the mean age and age spread, the rate of star formation, and circumstellar disks.
Newer results include those on the stellar/sub-stellar mass function 
as derived from deep near-infrared imaging 
(in collaboration with John Carpenter) and on the x-rays seen by 
the Chandra/ACIS instrument (in collaboration with Gordon Garmire, 
Eric Feigelson, and the rest of the Penn State Chandra/ACIS team).
I begin, however, with a summary of what is known globally about the clustering
of star formation in the Orion clouds.

\section{Clusters in the Orion Star-Forming Region}

It is not news that stars form in clusters, perhaps more frequently than not.
In fact, two of the more well-studied dense young clusters are located
in Orion -- the inner ONC (or Trapezium region) with a 
stellar density of 1-2$\times10^4$ stars/pc$^3$, in Orion A, and NGC 2024
with a stellar density of 1-2$\times10^3$ stars/pc$^3$, in Orion B.  
These clusters, and others like them outside of Orion, 
contain a mix of low- and high-mass stars.
They are the proof that stars of all masses (and brown dwarfs too) 
can and do form in the same place, within $<$0.05 pc, and at the same time, 
within $<$0.5 Myr.  That star formation occurs commonly in dense clusters means
we ought to consider the potential for mutual influence of cluster stars 
on each other and on their environments.  The relevant physical processes
include:

\begin{itemize}
\item
mergers/coagulation in the post-fragmention, proto-stellar stage;
\item
enhanced accretion in disks and/or scattering of planetessimals in 
post-accretion planet-building disks caused by gravitational interactions; 
\item
injection of mechanical energy into 
the ambient cloud by multiple misaligned outflows; 
\item
ionization of the cloud due to high levels of x-ray flux 
from protostars and young pre-main sequence stars; 
\item
massive star winds and ultraviolet radiation. 
\end{itemize}

Most nearby molecular clouds are in the process 
of forming stars -- at least in a global sense.  However, star formation 
is not happening at all places within these clouds at all times.  Rather,
it appears to occur in discrete places and at discrete times.
This is nicely illustrated in Orion A
by the chains of (presumably) protostellar/protocluster cloud cores 
in the northernmost part of the cloud (Chini et al., 1997; Johnstone \& Bally, 1999)
which are spatially distinct from the optically revealed 
ONC, which is in turn spatially distinct from the
pockets of clustered optical and still-embedded young stars located further
south (Carpenter, 2000; Strom, Strom, \& Merrill, 1993). 
Orion B displays similarly segregated behavior (Carpenter, 2000; Lada et al., 1991).  
Typical cluster sizes
are 0.2-0.5 pc while typical cluster densities are 100-200 stars/pc$^3$.
The full spectrum of cluster parameters remains unquantified: in these
same clouds exist the rich ONC and NGC 2024 clusters as mentioned above, but also
potentially hundreds of relatively poor, small aggregates which may have 
dispersed before we can detect them.  Only when it is possible to integrate
under a well-established distribution of cluster parameters can we begin to ask 
questions such as ``what is the probability that a star
of given mass is born in a cluster of given density.''  

Studies of both
the properties of the clusters and the properties of the stars which constitute
them are necessary in order to make progress in this area.  I turn now to 
discussion of some of these details in the ONC itself.

\section{Old Results on the Orion Nebula Cluster}

\begin{figure}[t]
\plotfiddle{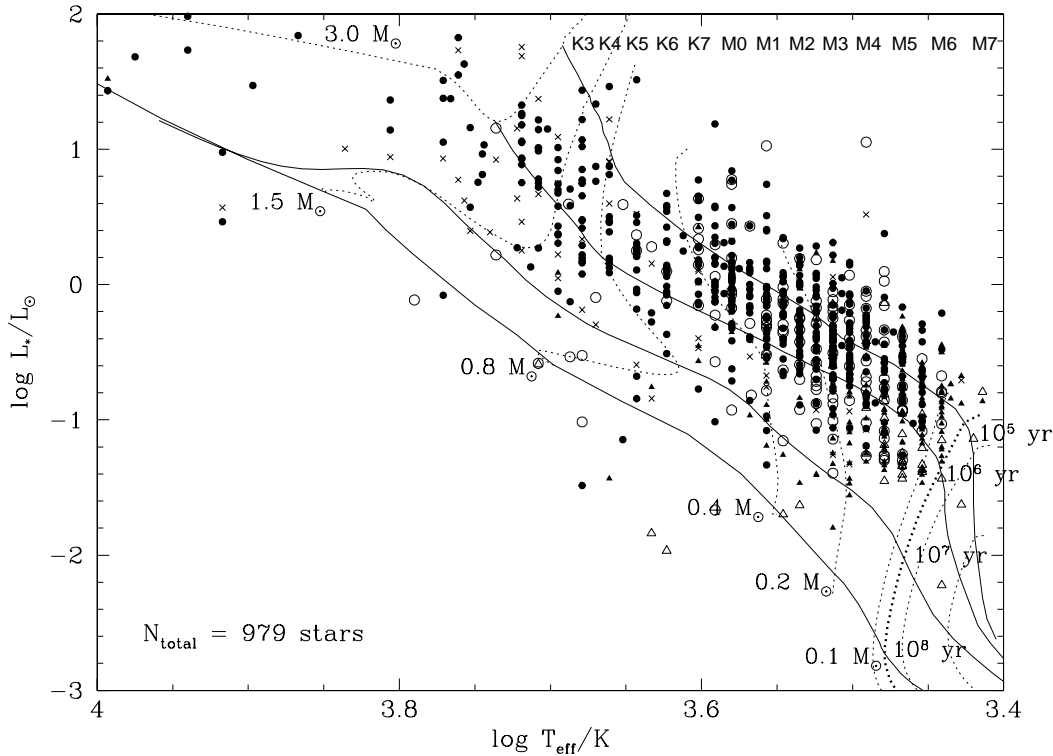}{3.in}{-90}{53}{53}{-210}{260}
\vskip 0.5truein
\caption{HR diagram for the Orion Nebula Cluster.
Triangles indicate lower limits in luminosity.
Filled circles/triangles indicate proper motion cluster members
plus sources identified as externally ionized;
open circles/triangles indicate that no proper motion information is available;
crosses indicate proper motion nonmembers.
Typical errors are $<0.02$ in log T$_{eff}$ and $<0.2$ in log L.
Superposed are the pre-main sequence evolutionary tracks of 
D'Antona \& Mazzitelli (1997, 1998).
\label{hrd}
}
\vskip -0.2truein
\end{figure}

Hillenbrand (1997) published a synthesis of census information combined with
new optical photometry and spectroscopy for stars within 2-3 pc of the center
of the ONC's Trapezium stars.  The HR diagram to a completeness
limit of I$_C$=17.5 for this sample is shown in Fig. 1, updated
to reflect current transformations from the observations to the HR diagram
and also more current theoretical tracks.  There remain serious discrepancies
at present between different sets of pre-main sequence evolutionary tracks.
I have shown the D'Antona \& Mazzitelli (1997, 1998) calculations largely
because they cover the wide range of effective temperatures and luminosities 
displayed by the data.  Other sets of tracks, e.g. those by 
Siess et al. (2000), Palla \& Stahler (1999), and Baraffe et al. (1998),
do not cover the full range of the ONC observations  
(see contributions to these proceedings by these authors for
discussion of detailed differences between tracks and the causal physics).

\begin{figure}[t]
\plotfiddle{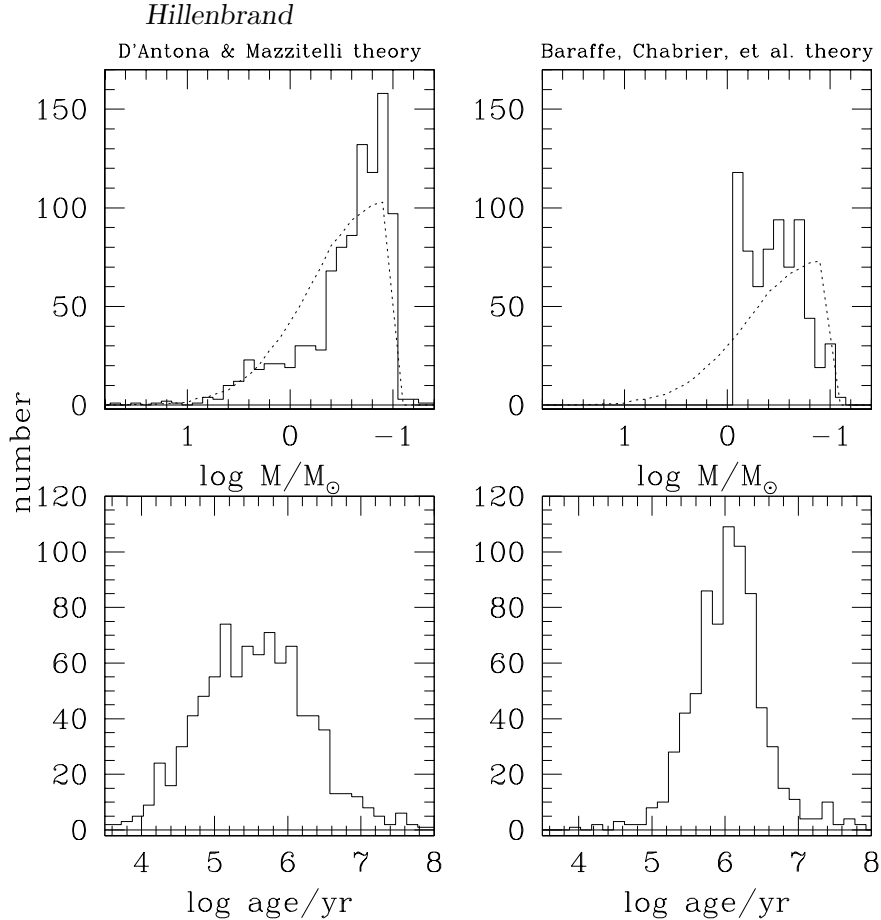}{2.in}{0}{60}{60}{-175}{-255}
\vskip 2.15truein
\caption{Mass and age distributions for the Orion Nebula Cluster.
Left panels show the results from D'Antona \& Mazzitelli (1997, 1998) while
right panels show the results from Baraffe et al. (1998).  Interpolations in
mass were restricted to those masses within the boundaries of the tracks while
interpolations in age were allowed to exceed the boundaries of the tracks.
A Miller-Scalo function truncated at 0.08 M$_\odot$ is shown for reference
in the top panels.
\label{deriv}
}
\vskip -0.29truein
\end{figure}

From the ONC data and the theory, one is able to derive quantities such as
the stellar mass spectrum and the star formation history.  We show in Fig. 2
the mass and age distributions produced by two different sets
of theory in order to illustrate their discrepancies.  In each of the age panels,
the distribution is gaussian in nature, a feature which suggests an increase
in the rate of star formation from past to present.  In each of the mass panels,
the distribution appears as falling towards the hydrogen burning 
limit, although this turnover nearly coincides with the completeness limit
of the data.  We pursue the shape of the initial mass function across the
hydrogen-burning limit in the next section.  From the combined age and mass
information, a recent star formation rate in excess of 10$^{-3}$ M$_\odot$/yr 
is implied.

\section{New Results on the Orion Nebula Cluster}

In this section I present more recently obtained results 
on the stellar/sub-stellar mass function in the inner ONC and
regarding the frequency with which x-ray emission 
is associated with known optical/infrared sources. 

\subsection{Extension of the Mass Spectrum into the Sub-Stellar Regime}

\begin{figure}[t]
\plotfiddle{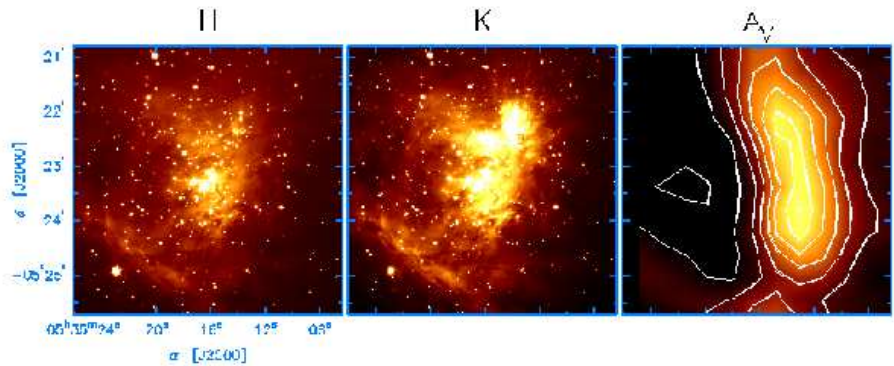}{1.in}{-90}{65}{65}{-245}{200}
\vskip 1.0truein
\caption{Images of our H and K-band mosaics from Keck/NIRC along with
an extinction map derived
from the molecular column density data of Goldsmith, Bergin, \& Lis (1997).
The pixel size of the infrared mosaics is 0.15$"$ and the angular
resolution of the extinction map is 50$"$. Contours in the
extinction map begin at A$_V$ = 5 mag and are spaced at
$\Delta$A$_V$ = 10 mag intervals.
}
\vskip -0.2truein
\label{khk_hess}
\end{figure}

Several groups have attempted recently to quantify the substellar mass
function in the inner ONC based exclusively on near-infrared data 
(Hillenbrand \& Carpenter, 2000; Luhman et al., 2000; Lucas \& Roche, 2000;
see also, Simon, Close, \& Beck, 1999).
The results of these studies appear quite similar in general terms, which is
especially encouraging given the gross differences in technique.

Our study was an imaging survey (see Fig. 3)
at K (2.20$\mu$m), H (1.65$\mu$m), and Z (1.05$\mu$m) 
covering $\sim$5.1'$\times$5.1'
centered on $\theta^1C$ Ori, the most massive star in the ONC.
For the age and distance of the cluster, and in the absence of extinction,
the hydrogen burning limit (0.08 M$_\odot$) occurs at
K$\approx$ 13.5 mag while an object of mass 0.02 M$_\odot$
has K$\approx$ 16.2 mag.  Our photometry is complete
for source detection at the 7$\sigma$ level to K$\approx$17.5 mag and
thus is sensitive to objects as low-mass as 0.02$M_\odot$ seen through
visual extinction values as high as 10 magnitudes.  We used the observed
magnitudes, colors, and star counts to constrain the shape of the inner ONC
stellar/substellar mass function.  To do so, we developed a new technique which
assumes the same stellar age and near-infrared excess properties which 
characterize optically visible stars in this same inner ONC region,
and extract a mass function based on probability analysis of the 
surface density of photometry in the K--(H-K) diagram.

\begin{figure}[t]
\plotfiddle{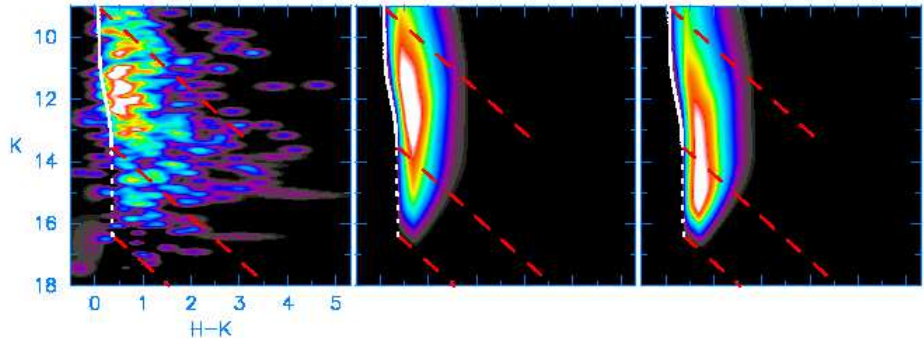}{1.75in}{-90}{60}{60}{-230}{260}
\caption{Simulations in Hess diagram format compared with
our near-infrared K vs H-K data. The models assume 1) an age distribution
which is log-uniform between 0.1 and 1.0 Myr, 2) a near-infrared excess 
distribution which is a half-gaussian in H-K and related linearly
to the monchromatic K excess, and 3) an extinction distribution which is uniform
in the interval A$_V$=0-5 mag.  The middle
panel shows the log-normal form of the Miller-Scalo mass function
while the right panel shows a shallow power law mass function
(N(log M) $\propto$ M$^{-0.35}$).  
Our field-star-subtracted data appear in the left panel. 
A falling mass function like that of Miller-Scalo better
represents the peak in the observed ONC star counts than does
an increasing mass function such as the shallow power-law.  
\label{khk}
}
\vskip -0.2truein
\end{figure}

\begin{figure}[t]
\plotfiddle{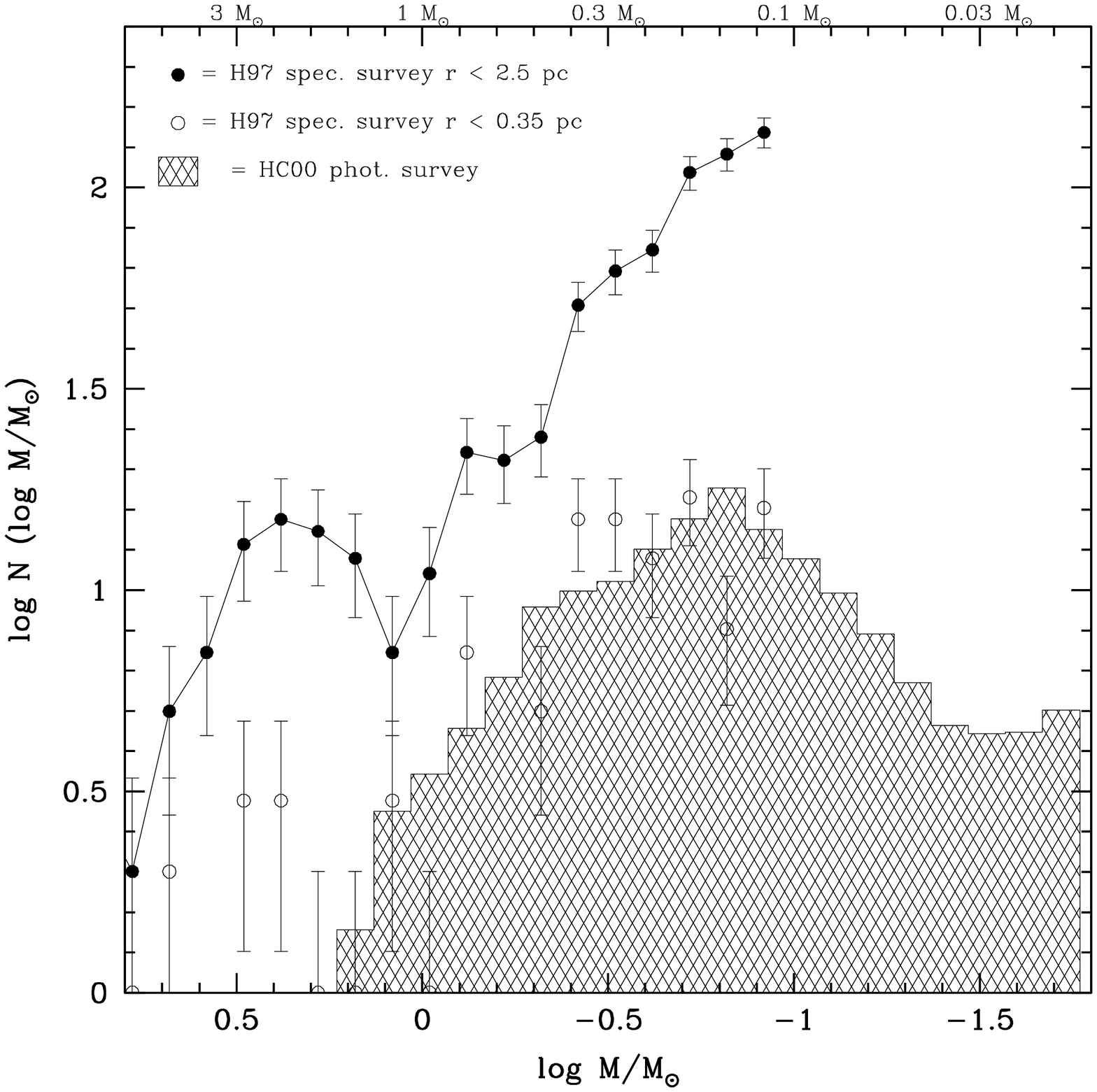}{2.in}{0}{50}{50}{-140}{-190}
\vskip 1.5truein
\caption{
Comparison of the ONC mass spectrum derived from optical
spectroscopic techniques (filled and open circles -- from data in Hillenbrand, 1997)
with that derived using infrared photometric techniques (histogram -- from data
and analyis of Hillenbrand \& Carpenter, 2000). 
No normalization has been
applied to these curves.  Note the general agreement
between the optical spectroscopic results and the near-infrared photometric
results in the mass completeness and the spatial area regimes where they
overlap (open circles vs hatched histogram).  Note also the disagreement
between the shape of the mass spectrum derived for the inner ONC (r$<$0.35 pc;
open circles) vs the greater ONC (r$<$2.5 pc; filled circles).
\label{imf}
}
\vskip -0.25truein
\end{figure}

We find that our data are inconsistent with a mass function that rises
across the stellar/sub-stellar boundary, as shown in Fig. 4.  
Instead, we find that the most likely
form of the inner ONC mass function is one that rises to a peak around
0.15 M$_\odot$, and then declines across the hydrogen-burning limit
with slope N(log M) $\propto$ M$^{0.57\pm0.05}$, as shown in Fig. 5.
Our conclusions for the substellar mass function apply to the inner 
0.71 pc x 0.71 pc of the ONC only; they may not apply to the ONC 
as a whole where some evidence for general mass segregation has been found
(Hillenbrand \& Hartmann, 1998).

\subsection{X-rays from Chandra}

\begin{figure}[t]
\plotfiddle{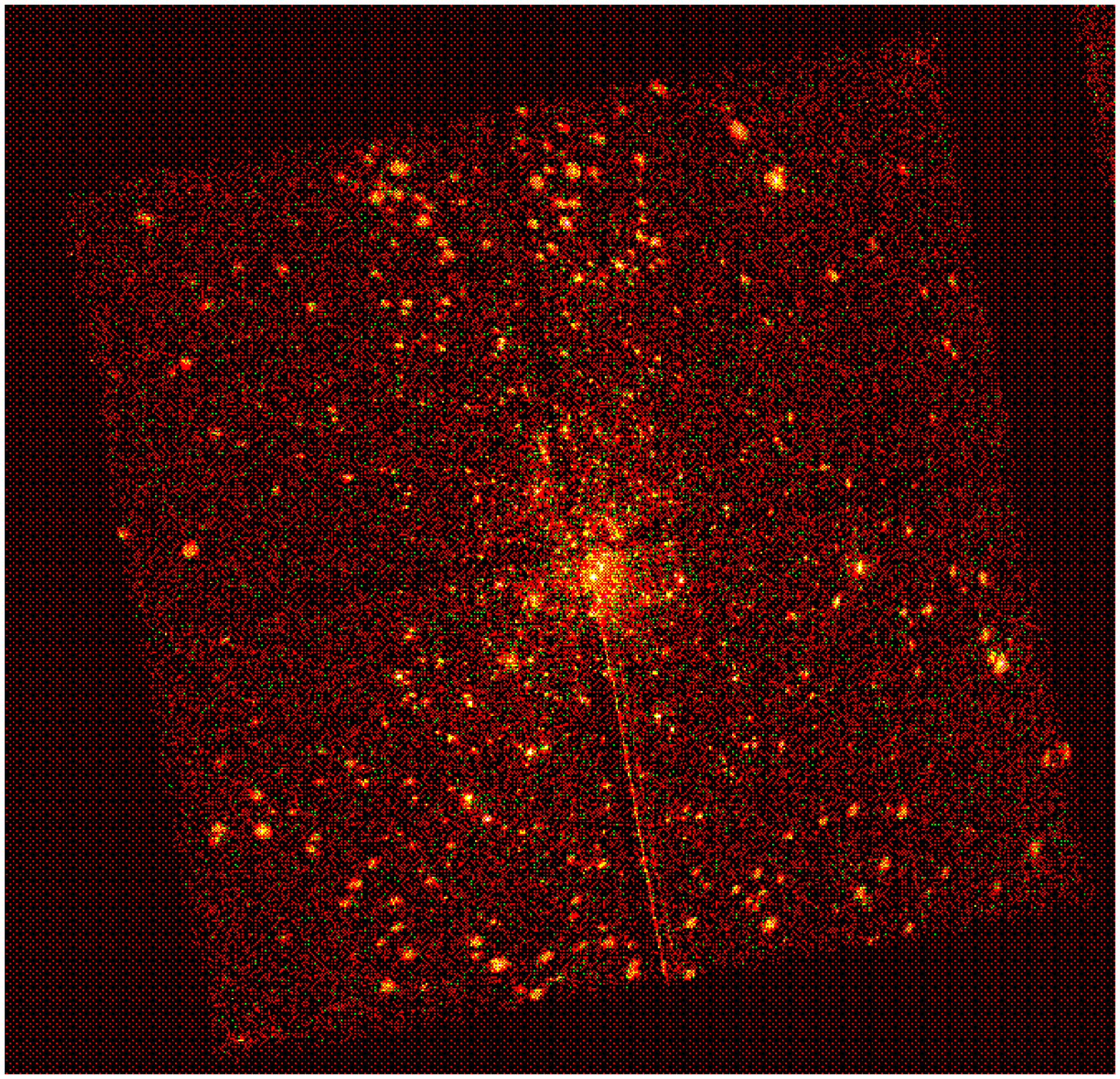}{2.in}{0}{70}{70}{-210}{-290}
\vskip 2.75 truein
\caption{ACIS-I image of the Orion Nebula Cluster in the
$0.2-8$ keV band showing $\simeq 1000$ X-ray sources.  The
17\arcmin$\times$17\arcmin\/ array is shown here at a reduced resolution
of 2\arcsec$\times$2\arcsec\/ pixels.  Intensity scaling   
is logarithmic according to the number of events. 
North is up and East is to left.
\label{acis}
}
\vskip -0.35truein
\end{figure}

Pre-main sequence stars are known to display levels of x-ray emission which
are 1 to $>$4 orders of magnitude higher than their counterparts on
the main sequence.  The x-rays are variable, exhibiting flaring characteristics 
similar to those of the active sun, and are attributed to $\sim10^6$ K plasma
heated by magnetic reconnection events 
(see review by Feigelson \& Montmerle, 1999).

The Orion region is an old and familiar target of x-ray satellites. The inner 
ONC region was recently observed during GTO programs with
several of the instruments on Chandra. 
I will describe preliminary results from the Chandra/ACIS (Advanced CCD Imaging
Spectrometer) instrument team; see the contribution 
of Harnden for preliminary results from the Chandra/HRC instrument team.
The exquisite spatial resolution ($\sim$0.5'') and wide field of view 
of ACIS combined with its low background make it unprecedented for x-ray studies
of crowded regions such as the ONC.  In Fig. 6 we show the 
17 x 17 arcmin$^2$ ACIS GTO image (48 ksec exposure) of Garmire et al. (2000), 
claimed to be the 
richest astronomical x-ray image yet obtained with close to 1000 point sources.  

Preliminary results based on these data are described in Garmire et al. (2000)
with a more detailed paper in preparation (Feigelson et al., 2000).  The 
sensitivity at a limit of 7 photons in the 0.2-8 keV band is to luminosity
$\sim2\times10^{28}$ erg/s assuming kT = 1 keV and little obscuration. 
A total of 831 sources above this completeness limit and 142 below it have
been identified.
Thusfar we have cross-correlated the list of ACIS detections with optical
and near-infrared sources in the lists of Hillenbrand (1997), Hillenbrand et al.
(1998), and Hillenbrand \& Carpenter (2000).  The optical surveys are complete
over the full area of the ACIS image to V$\approx$ 20 mag and several magnitudes
deeper in the inner 3 x 3 arcmin$^2$ based on the work of Prosser et al. (1994).
The near-infrared surveys are complete over the full area of the ACIS image to 
K$\approx$ 13.5 mag and several magnitudes deeper in the inner 5 x 5 arcmin$^2$. 
Of the 973 x-ray sources, 860 coincide within $<$1 arcsec ($<$2 arcsec in
the outermost portions of the x-ray image) with an optical/infrared star.
Others are associated with compact radio sources (Menten et al., private
communication; Felli et al, 1993; Churchwell et al., 1987) or so-called
ProPlyDs (e.g. O'Dell \& Wong, 1996) without stellar point sources.  Objects of all
masses ranging from the 50 M$_\odot$ star $\theta^1$C Ori to those at or just
below the 0.08 M$_\odot$ hydrogen burning limit are detected.  As a reminder,
the stars in this region are $<$1-3 Myr of age.

Comparison with an HR diagram complete in mass down to $\sim$0.1 M$_\odot$ for 
extinction A$_V<$ 2 mag shows
that Chandra/ACIS detected 91\% of stars with M/M$_\odot >$ 0.3 and 75\% of
those with 0.1 $< M/M_\odot <$ 0.3.  Considering the deep near-infrared survey
discussed above which is complete to  M/M$_\odot >$ 0.02 for A$_V<$ 10 mag,
most of the {\it stars} are detected although those with higher extinction
generally are not detected and neither are substellar mass objects
at any extinction.
It is not known at present whether the lack of detection of candidate brown
dwarf objects reflects a true astrophysical phenomenon or an L$_X$ vs L$_{bol}$
or L$_X$ vs M relationship combined with the Chandra/ACIS limits.

Because Orion is a well-studied region, estimates for physical variables such
as stellar masses, ages, radii, and rotational velocities exist for many 
hundreds of stars.  Combination of these stellar parameters (and perhaps
even circumstellar parameters such as disk accretion rates) with the Chandra 
x-ray data should enable sorting between various possibilities for the 
currently confounding genesis of x-ray activity in pre-main sequence stars.
Thusfar our analysis has suggested a relatively constant level of L$_X$ at
early stages followed by development at older ages (corresponding to 
lower luminosities as the stars contract) of a large dispersion in x-ray 
activity for stars of the same mass and age.

\begin{figure}[t]
\plotfiddle{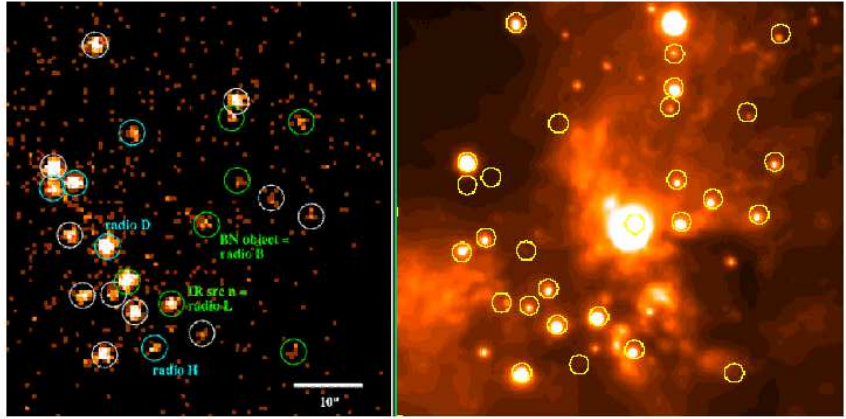}{2.2in}{-90}{50}{50}{-190}{245}
\vskip -0.1truein
\caption{
Comparison of x-ray and near-infrared images of the BN/KL region.
The ACIS image is from 2-8 keV while the NIRC image is at 2$\mu$m.
Circles in the left panel indicate hard x-ray sources while circles in the
right panel indicate either hard or soft x-ray sources.
Images are $\sim$1 x 1 arcmin$^2$ in size and centered on the BN object.  
\label{acisbnkl}
}
\vskip -0.25truein
\end{figure}

Chandra/ACIS also gives us a clearer and deeper view into
the BN/KL region which is embedded in molecular gas just behind the ONC, 
a site where massive star formation is currently taking place.  
In Fig. 7 we show a close-up of Fig. 6
comparing the ACIS-detected x-ray sources with the Keck/NIRC near-infrared 
sources.  While the majority of
x-ray sources over the full field of view of ACIS do have optical/infrared
counterparts, the situation is slightly different in the BN/KL region.
Source n is detected with ACIS, but the identification of the BN object with
its closest X-ray neighbor ($\sim 0.3"$, i.e., $3 \sigma$ separation)  is
unclear at the moment. The X-ray source may rather be associated with the BN
extended circumstellar environment. IrC2, the third powerful embedded massive
prototellar object in the region, is not detected.  Several hard x-ray sources
in the left panel do not have infrared counterparts in the right panel; 
most of these have been detected in the radio.  For contrast,
there are also a number of sources which are bright at K-band but which lack
x-ray counterparts.  

\section{Concluding Remarks}

Despite more than six decades of study of the Orion Nebula Cluster,
we continue to learn new things about it with every improvement in technology. 
In some sense the Orion star-forming region is a prototype: it is the nearest 
Giant Molecular Cloud and the nearest region of recent massive star formation.  
As such, it is considered an important laboratory for testing our 
understanding of star formation processes in the Galaxy.  Equally important,
however, is discovering how typical Orion-like regions are in the Galaxy.

\acknowledgements
I would like to acknowledge my collaborators on these projects: Carpenter, 
Feigelson, Broos, Garmire, Pravdo, Townsley, and Tsuboi, without whom 
this presentation would not have been possible.  

\end{document}